\documentclass[aps,amsmath,amssymb,twocolumn,pre,nofootinbib]{revtex4-1}

\usepackage{graphicx}
\usepackage{color}
\usepackage{bbold}

\usepackage[usenames,dvipsnames]{xcolor}
\usepackage{amsmath}
\usepackage{amsthm, amssymb}
\usepackage{enumitem}
\usepackage{slashed}
\usepackage{tikz}
\usetikzlibrary{calc,fadings,decorations.pathreplacing,shapes,shapes.multipart,arrows,shapes.misc,intersections,positioning,patterns}
\usepackage{bm}
\usepackage{dsfont}
\usepackage{changepage}
\usepackage{array}

\definecolor{bluepurple2}{rgb}{0.06,0,0.6}
\usepackage[colorlinks=true,citecolor=blue,linkcolor=bluepurple2]{hyperref}

\usepackage{bm}

\graphicspath{{./}{./images/}}

\newcommand{\bit}{\begin{itemize}}
\newcommand{\eit}{\end{itemize}}

\newcommand{\ba}{\begin{align}}
\newcommand{\ea}{\end{align}}
\newcommand{\be}{\begin{equation}}
\newcommand{\ee}{\end{equation}}
\newcommand{\bi}{\begin{itemize}}
\newcommand{\ei}{\end{itemize}}

\newcommand{\M}{\mathcal{M}}
\newcommand{\RP}{\mathrm{RP}}
\newcommand{\CP}{\mathrm{CP}}

\begin{document}

\title{3D Unoriented loop models and the $\RP^{n-1}$ sigma model}

\author{Pablo Serna}

\affiliation{Departamento de Física - CIOyN, Universidad de Murcia, Murcia 30.071, Spain}

\date{\today}

\begin{abstract}

We study a completely-packed loop model with crossings in a three-dimensional lattice and confirm it is described by $\RP^{n-1}$ sigma field theories. We use Monte Carlo simulations, with systems sizes up to $1800\times1800\times1800$, to obtain the critical exponents for a universality class with no previously reported estimates in the literature, namely the replica-like limit $n\rightarrow1$ of $\RP^{n-1}$ sigma field theory. Estimates of critical exponents include $\nu=0.918(5)$ and $\eta=-0.091(9)$. We also study the scaling dimension of the 4-leg watermelon correlators, particularly for $n=1$ we obtain the value $x_4=1.292(8)$. 

\noindent
\end{abstract}
\maketitle

\suppressfloats

\section{Introduction}

The relationship between geometrical problems, such as percolation, models of membranes or loop models, and quantum and classical statistical mechanics problems is fascinating and sometimes elusive. 
Classical loop models, in particular, form an ensemble of problems interesting by themselves and with a wide range of phenomena \cite{deGennes1972,duplantier1987exact, affleck1991nonlinear, ardonne2004topological,read2001exact, jacobsen2009polygons, nahum2012universal}.
Relevant to this work, they can be related to problems of polymers \cite{cardy1999logarithmic,candu2010universality,nahum2013loop,nahum2016universality}, Anderson transitions \cite{beamond2002quantum,ortuno2009random} and quantum magnets \cite{nahum2013phase, nahum2015deconfined}, among many others.

In \cite{nahum2013loop}, it was studied a two-dimensional completely-packed loop model with crossings (CPLC), whose continuous transitions are described by a sigma theory for a field on the real projective space $\RP^{n-1}$ \cite{de1993physics,nahum2012universal,nahum2013loop}. In the replica limit of $n\rightarrow1$, it was found a striking similarity with the critical exponents of the symplectic class of metal-insulator
Anderson transitions \cite{evers2008anderson}. Despite the differences between the two systems, a simple argument
seems to relate the loop model in two dimensions with a quantum network 
model. This is analog of the map between the class C of Anderson transitions and a network model
\cite{beamond2002quantum}, 
which holds in any dimension,
in particular in 3D \cite{ortuno2009random}, as long as the coordination number of the lattice is 4. By analogy, it raises the question of whether the generalization to higher spatial dimensions holds in the CPLC or not.

In the loop representation of $\RP^{n-1}$, $n$ is a fugacity to the number of loops. 
This sigma field theory describes many interesting problems and it has been studied thoroughly in two spatial dimensions \cite{caracciolo1993new,hasenbusch1996n,niedermayer1996question,bonati2020asymptotic}, specifically, for $n<2$ there is a continuous order-disorder transition in 2D \cite{nahum2013loop}.
In three dimensions it is interesting too as it describes, for example, the well-known $O(2)$ model or XY, for $n=2$ \cite{campostrini2001critical, nahum2013loop, nahum2013phase}. 
For values of $n=3$, it describes the nematic to isotropic phase transition in a liquid crystal, the so-called Lebwohl-Lasher model \cite{lebwohl1972nematic, duane1981phase}. 
Note that this phase transition is weakly first-order, and there are many studies characterizing the associated liquid crystal system \cite{fabbri1986monte,priezjev2001cluster,jayasri2005wang}.
As suggested in \cite{nahum2013phase}, the character of the transition may be due to the annihilation of a tricritical and a critical point, in a lower value of $2<n<3$. We will study this in more detail in a future work \cite{n3_weaklyfirstorder}.
The literature for the replica-like limit $n\rightarrow1$ in three dimensions, on the other hand, is scarce  \cite{duane1981phase, kunz1989first,nahum2012universal}. 
As in other similar sigma models, where this limit is related to polymers, percolation, and other geometrical systems 
\cite{parisi1980self,mckane1980reformulation,cardy1999logarithmic,dubail2010conformal}, the partition function is zero but the problem remains non-trivial. As a side note, the literature on the related \emph{antiferromagnetic} $\RP^{n-1}$ models is abundant and they show different universality classes \cite{fernandez2005numerical,pelissetto2018criticality,reehorst2020bootstrapping}.

Loop models also give access to different kind of operators that are less common in classical stat-mech problems. 
Particularly, watermelon correlators \cite{saleur1987exact,duplantier1989two,cardy2005sle,janke2005fractal,jacobsen2009polygons} are important due to their relationship to anisotropy terms \cite{hasenbusch2011anisotropic} and their quantum counterpart \cite{dai2020quantum}. 
In 2D systems where conformal field theory works, there is a good understanding of the behaviour of these operators and their scaling dimension in some models. 
For the $\RP^{n-1}$, numerical estimates have been obtained in 2D \cite{nahum2013loop}. 
However, in 3D classical systems or 2+1D quantum systems (and 3D disordered electrons), the estimates come either from the scaling dimensions of anisotropic perturbations \cite{hasenbusch2011anisotropic} or from conformal bootstrap of rank-2 tensors \cite{chester2020carving,reehorst2020bootstrapping}. 

We describe here the CPLC model in a three-dimensional lattice and characterize the critical exponents governing its phase transition lines. 
Monte Carlo calculations show that it harbours a new universality class that has not been described before in the literature to the best of our knowledge.  
This universality class does not seem to be related to the universality class of symplectic Anderson transitions in 3d  (nor any of the usual classes). 
We study the critical parameters at two different points on the critical line with compatible results for values of $n=1$. 
Also, we analyze the model and its phase diagram for integer fugacities $n>1$. Finally, we provide estimates of the scaling dimensions of 2 and 4 leg watermelon correlators for several points in the CPLC.

\section{Model}\label{model}
The model we study is a natural extension to three dimensions 
of the two-dimensional CPLC defined in the square lattice 
\cite{nahum2013loop}. We use the three-dimensional L-lattice proposed by 
Cardy \cite{cardy201050}, stripping the links of their fixed orientation, but keeping a coordination number 4, as in the square lattice. We will review briefly in this section both the definition of this lattice and the CPLC. 

\def\Z{\mathbb{Z}}

\begin{figure}[ht]
 \includegraphics[width=0.48\linewidth]{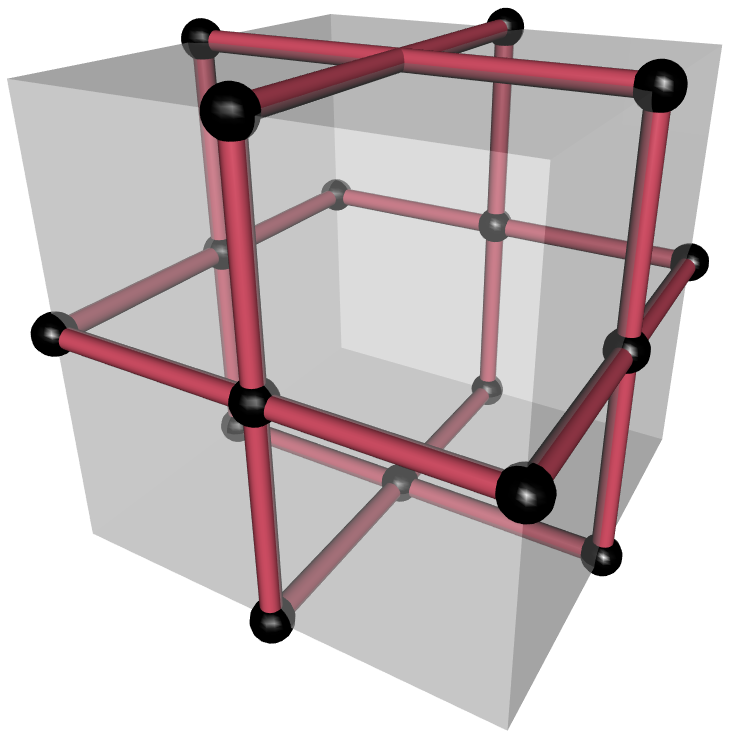}
 \includegraphics[width=0.48\linewidth]{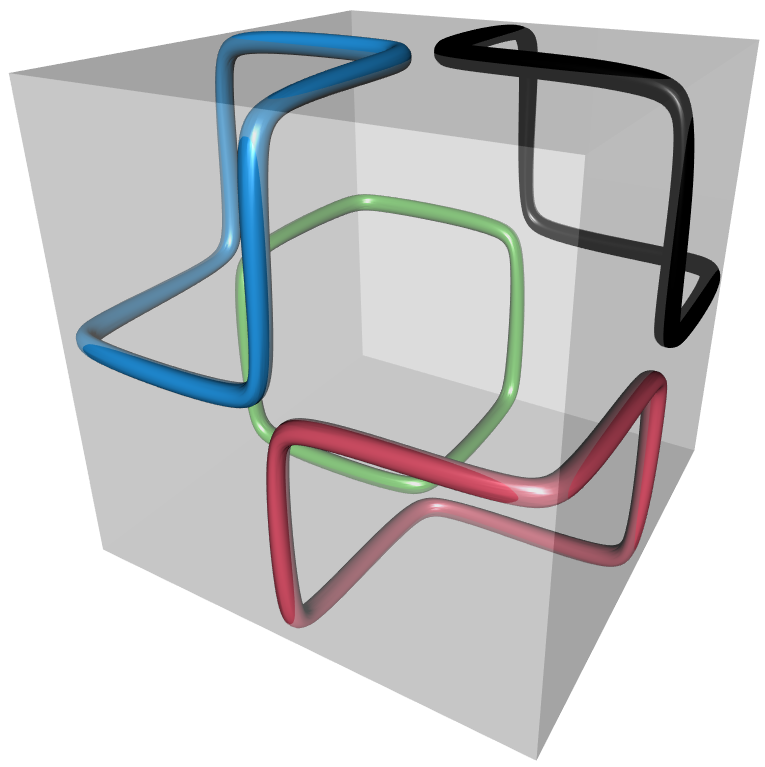}

 \caption{Left panel: Structure of the lattice where loops are defined, each node can pair its links in three ways. Right panel: Loop sample with $p=0$ and $q=0$, the cube repeats itself with fcc symmetry.}
 \label{fig:lattice}
\end{figure}

\begin{figure}[ht]
\includegraphics[width=\linewidth]{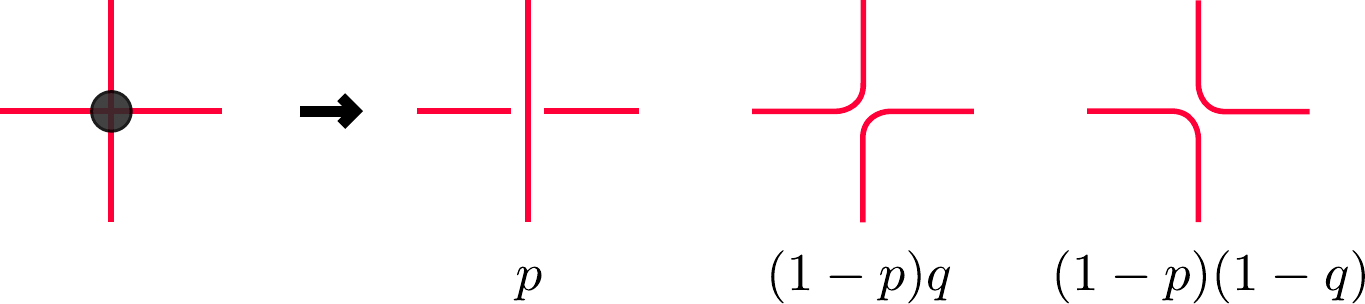}
\caption{Three possible pairings of the links with their corresponding weights.}
\label{fig:pairlinks}
\end{figure}

\begin{figure}
  \includegraphics[width=0.49\linewidth]{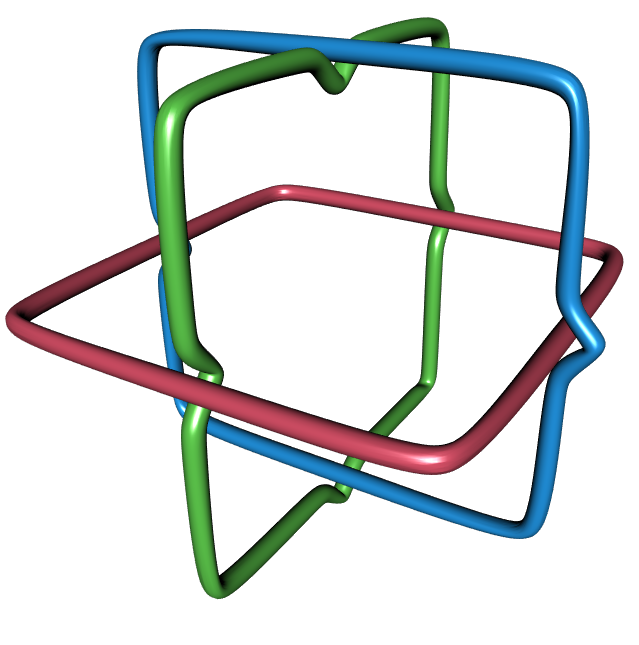}
  \includegraphics[width=0.49\linewidth]{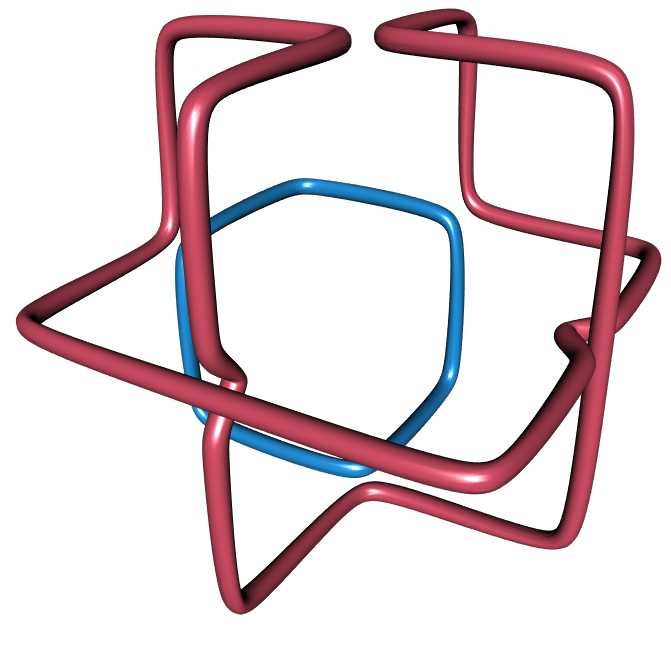}
\caption{Two configuration examples with crossings and short loops.}\label{fig:examples}
\end{figure}

In the 3d L-lattice,  the links are formed by the intersection of the faces of the two lattices $(2\Z)^3$ and $(2\Z+1)^3$, see Fig.~\ref{fig:lattice}. The four links are coplanar for each node and belong to the faces of a simple cubic lattice. 
Then, a loop configuration is constructed by pairing these four links at each node in three possible configurations illustrated in Fig.~\ref{fig:pairlinks}. 
The statistical weight of such a configuration has two contributions. 
First, the product of the weight associated with each pairing: $p$ 
when opposite links are paired and there is a cross, while $(1-p)q$ 
and $(1-p)(1-q)$ are associated with the other two possibilities. 
When there is a crossing, the weight is not affected by which links go ``out of the plane''. A crossing may join strands of two separate loops or of a single loop, as in the examples shown in Fig.~\ref{fig:examples}.  
Second, each loop may carry a fugacity $n$, which we use as $n$ possible colorings of the loops. The Boltzmann factor of a given configuration is then
\begin{equation}
 p^{n_p}[(1-p)q]^{n_q}[(1-p)(1-q)]^{n_{1-q}} n^\text{n. of loops}\quad,
\end{equation}
where $n_p$, $n_q$ and $n_{1-q}$ are the numbers of plaquettes with the given weights. Then, the partition function is the sum of this factor for all given configurations with $n_p+n_q+n_{1-q}=N$ fixed by the lattice.

Now, the model is fully specified by the choice of which of the non-crossing pairings wear which weight at each node. We specify the only surviving configuration at $q=0$ and $p=0$, Fig.~\ref{fig:lattice} right panel. This sets unambiguously the weights $(1-p)q$ and $(1-p)(1-q)$. This configuration has four kinds of loops of length six and an fcc symmetry. We can describe them by giving a starting point and three first steps, the next three final steps are the opposite in the same order, see table~\ref{hexagons}. We use in all simulations periodic boundary conditions. Fig.~\ref{fig:examples} shows two examples of small loops with crossings, and Fig.~\ref{fig:samples long loops} in appendix \ref{app:samples} shows an example of a larger fractal loop. 


\begin{table}[htbp]
\centering
\begin{tabular}{crrr}
\hline
\hline
\multicolumn{ 4}{c}{L-Lattice} \\ \hline
$(0,1,0)$ & $\hat{\bf x}$&$\hat{\bf y}$&$\hat{\bf z}$\\ \hline
$(1,0,1)$ & $\hat{\bf x}$&$\hat{\bf y}$&$-\hat{\bf z}$\\ \hline
$(0,1,2)$ & $\hat{\bf x}$&$-\hat{\bf y}$&$-\hat{\bf z}$\\ \hline
$(1,2,1)$ & $\hat{\bf x}$&$-\hat{\bf y}$&$\hat{\bf z}$\\ \hline\hline
\end{tabular}
\caption{Initial position and first three steps of the hexagons forming the L-lattice.}
\label{hexagons}
\end{table}

\section{Phase diagram}

\begin{figure}[ht]
 \includegraphics[width=0.9\linewidth]{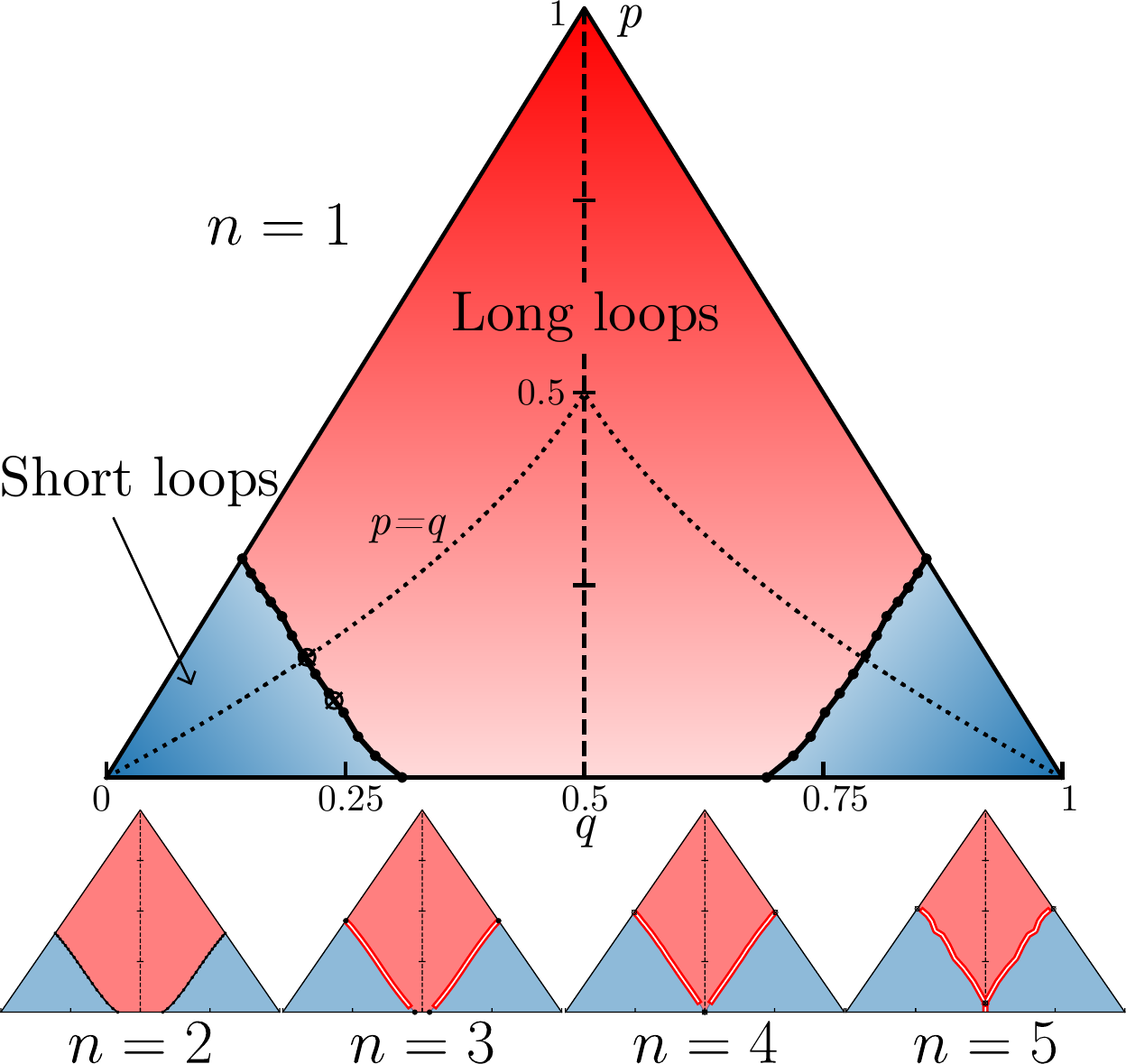}
 \caption{Phase diagram of the 3D CPLC for $n=$1, 2, 3, 4 and 5. Main panel shows the phase diagram for $n=1$. The two studied critical points at $p=q=0.156187$ and at $(p,q) = (0.1, 0.209147)$ are shown as big circles. The solid black lines represent continuous phase transitions, while red and white lines are first-order transitions.}
 \label{fig:phasediagram}
\end{figure}

The phase diagram of this loop model depends on the number of colors $n$, as can be seen in Fig.~\ref{fig:phasediagram}. In here, a line of constant value of $q$ is a straight line connecting $q$ at $p=0$ to the apex.
At $n=1$, we have mapped the phase diagram using system sizes $L=100$ and $200$, by fixing different values of $q$ and varying $p$. In Sec.~\ref{sec:pq} and \ref{sec:p01}, we provide a more careful study of the transition along the line of $p=q$ (dotted line) and the line of constant $p=0.1$.
When $n$ increases, the phases with short loops (blue) span a larger region of the phase diagram. 
For a given $n>n^*$, with $4<n^*<5$, the phase with long loops dissapears at the lower boundary (note that for $n=4$ the transitions at $p=0$ occur at $q$ and $1-q\approx 0.4994$ \cite{nahum2013phase}). 
The boundary line between phases in the interior of the phase diagram changes from continuous to first-order at some value of $2<n<3$. 
Some parts of the phase diagram have been studied in previous works and are related to already well-known models.
For an expanded version of the phase diagrams for $n\ge2$, see Fig.~\ref{fig:phdiagramsalln}. We provide here a more detailed description, reviewing what is already known.
 
 {\bf Boundaries --} On the lines $p=0$, $q=0$ or $q=1$ the model has a higher symmetry. Now the links can be given a fixed orientation without changing the partition function. The appropriate field theory is a sigma model for a field on the complex projective space $\CP^{n-1}$ \cite{nahum2013loop}. 
 In particular, $p=0$ corresponds to the phase diagram in the L-lattice for the completely packed loop model \cite{nahum2013phase}. 
 In the replica limit, $n\to1$, the model maps to a quantum network model of the class C of Anderson transitions \cite{beamond2002quantum,ortuno2009random}. 
 When $1<n<4$ the phase transition is in the universality class of the $\CP^{n-1}$ sigma field theories or $SU(n)$ quantum magnets in 2+1d \cite{nahum2013phase}. 
 For $n\ge n_c\ge3.0(2)$ the phase transition becomes first-order and in particular when $n>n^*$ with $4<n^*<5$ the phase with long loops disapears. 
 For the $q=0$ and $q=1$ lines, the behaviour is similar to the line $p=0$, but there is always a phase with long loops. A similar situation happens in an alternative lattice, K-lattice, defined in \cite{nahum2013loop}.
 
 {\bf Long loop phase --} The long loop phase region corresponds to a phase with Brownian loops, whose fractal dimension is 2. 
 The statistics of this phase is determined by the statistics of the ordered phase of the $\RP^{n-1}$ sigma model. 
 With the appropriate boundary conditions,  in the interior of the phase diagram the probability distribution of the length of the loops is a Poisson-Dirichlet with parameter $\theta=n/2$, while on the boundaries the distribution follows a parameter $\theta=n$ \cite{nahum2013length}.

 {\bf Critical line --} The lines that separate the long loop phase and short loop phases can either be continuous or first-order depending on the value of $n$. 
 When continuous, these critical lines correspond to phase transitions described by $\RP^{n-1}$ sigma field theory, driven by the proliferation of lines of $\Z_2$ defects (or $\Z$ defects for $n=2$), except  when the line reaches the boundary of the phase diagram. 
 As far as we know the replica-like limit $n\to1$ of this field theory in three dimensions has not been described previously in the literature. 
 
The long-distance behaviour close to the critical point is described by.
\begin{equation}
\mathcal{L} = \frac{K}{2}\text{tr}(\nabla Q)^2\quad,
\end{equation}
with $Q$ a real symmetric traceless $n\times n$ matrix under the constrain $(Q+1)^2=n(Q+1)$ \cite{nahum2013loop, nahum2013phase}. 
Note that we can use a redundant formulation in terms of real spins $s$ with $n$ components and the constraint $|s|^2=1$, by defining $Q^{ab} = s^as^b-1$. 
In this alternative representation, spin operators can be defined $S^a$ and this will be helpful to define higher-order correlators \cite{nahum2013loop}. As we said,
for $n=2$ this theory describes an $O(2)$ model or the three-dimensional XY model. 
For $n=3$, it describes a liquid crystal isotropic-nematic transition and is weakly first-order \cite{lebwohl1972nematic}. 

 {\bf Line $\mathbf{q=1/2}$ --} When $n> n^*$, there is a small but finite region, $0\le p<p_c$, where the two short loop phases are separated by a first-order transition at $q=1/2$ 
 (Fig.~\ref{fig:phasediagram}). This first-order transition ends at the intersection of two phase transition boundary lines, $p_c\approx 0.045$, raising the possibility of a different kind of critical point, either multicritical or a critical point where the topological defects of the field theory are suppressed. The latter option is akin to the deconfined criticality scenario \cite{senthil2004deconfined,wang2017deconfined}.
 Indeed, when $q=1/2$, the lattice has a higher symmetry \cite{nahum2015deconfined}, and with the parameter $p$ the  system can be driven from a short loop phase with four degenerated short loop states (analogous to a valence bond solid) to a phase with long loops. 
 Also at this line, it is natural to expect that the fugacity associated with topological defects of the $\RP^{n-1}$ model, lines  $\Z_2$ vortices (or $\Z$ for $n=2$) get suppressed, similarly to what it happens in the two-dimensional L-lattice \cite{nahum2013loop} and at the boundary $p=0$ \cite{nahum2015deconfined}. 
 This perturbation can be relevant and change the universality class of the critical point, compared to the phase transition found on the boundary lines. 
 However, results for $n=5$ show that this transition is first-order as we discuss in Sec.~\ref{sec:ngt1}. The most likely scenario is that this holds for larger $n$. 

\section{Phase transitions} \label{critical line}
The critical line of this loop model for $n=1$ presents a new universality class that has not been studied in detail before in the literature. We have estimated critical parameters at two different points of the critical boundary line.

\subsection{Line $p=q$}
\label{sec:pq}
\begin{figure}[ht]
 \includegraphics[width=\linewidth]{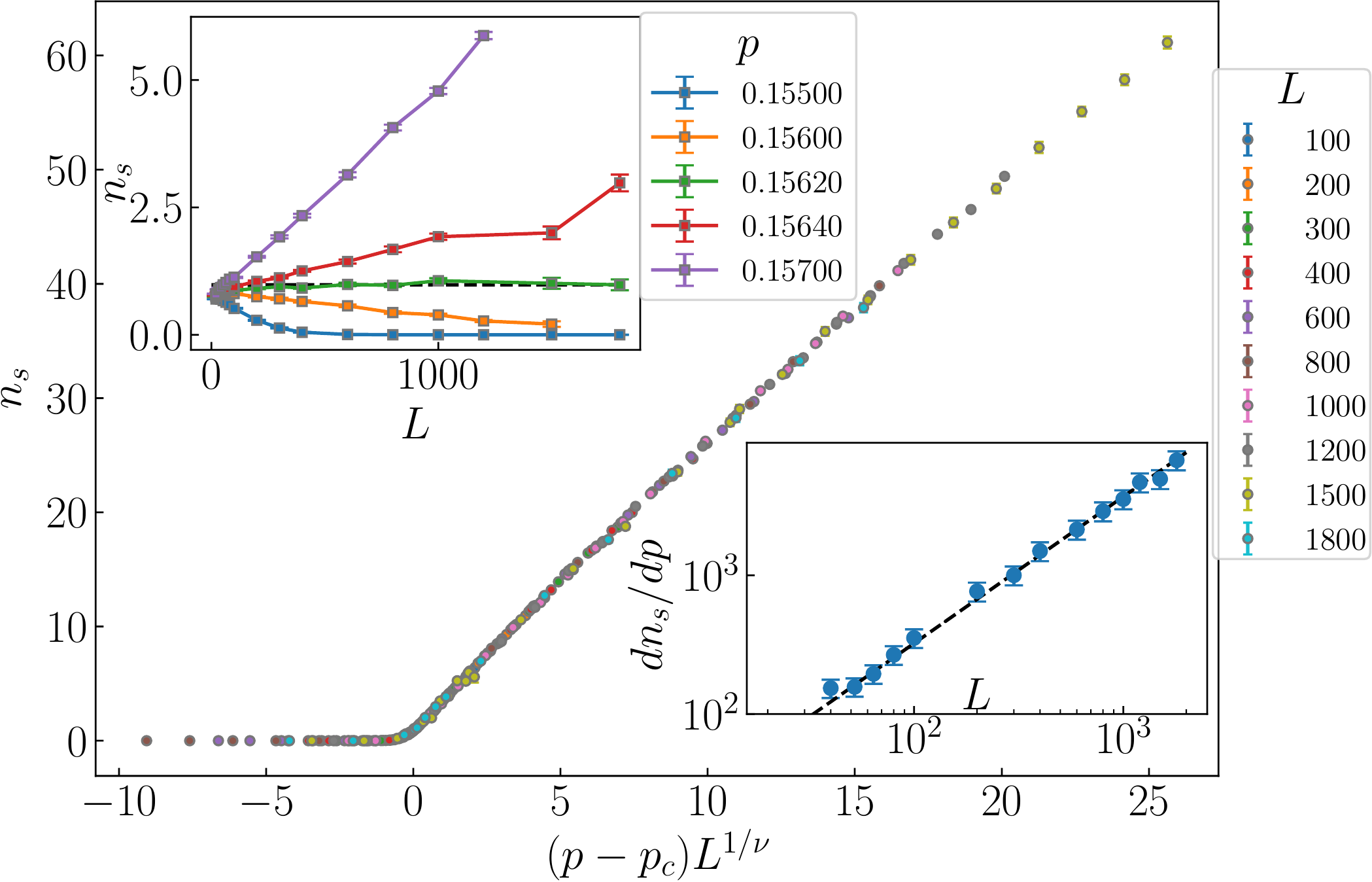}
 \caption{$\mathbf{n=1,\,p=q}$. Scaling collapse of $n_s$ as a function of $(p-p_c)L^{1/\nu}$ for the line $p=q$. Upper inset: spanning number as a function of $L$ for several values of $p$ close to $p_c=0.1562$. Lower inset: Slope of $n_s$ at the critical point as function of the system size. Black dashed line represents a power-law fit, $L^{1/\nu}$, to system sizes $L>100$, with value $\nu=0.945$.}
 \label{fig:ns}
\end{figure}

First, we have studied the loop model along the line $p=q$. The critical point can be found with very good accuracy by looking at the crossings of the number of strands that span the sample, spanning number or $n_s$, or its behaviour as a function of the system size, $L$.
\footnote{More precisely, $n_s$ is defined by cutting the sample by the plane $z=0$ and counting the number of strands that come out of this plane and reach the top $z=L$.} 
This quantity is related to the \emph{stiffness} and to similar topological quantities such as the number of winding loops.
In the phase with short loops, $n_s$ decreases exponentially with $L$, while in the long loop phase it grows linearly with $L$, see upper inset of Fig.~\ref{fig:ns}. 
At the critical point, $n_s$ tends to a critical value $n_s^*$ that only depends on the universality class. 
This flow of $n_s$ with the system size gives us a first estimate of the critical point $p_{\rm c}=0.15620(5)$ and the critical value $n_s^*\approx0.97$. A similar analysis using the crossings between curves of consecutive system sizes gives similar and compatible results. 
Close to the critical point, $n_s$ follows a scaling law of the form $n_s=f_{n_s}(L^{1/\nu}(p-p_{\rm c}))$, neglecting irrelevant exponents. A first estimate of the critical exponent $\nu$ can be obtained from the derivative of $n_s$ at $p_c$. This quantity grows with the system size as $d n_s(p_c)/dp \propto L^{1/\nu}$. 
In the lower inset of Fig.~\ref{fig:ns}, we show this quantity as a function of $L$. A fit to a power-law $L^{1/\nu}$ (black dashed line), restricted to sizes $L>200$, provides a rough estimate $\nu\approx 0.95$. 

\label{transition}

These two parameters  can  be extracted by obtaining a scaling collapse of the curves as it is shown in the main panel of Fig.~\ref{fig:ns}. For the construction of the function $f_{n_s}$, we use cubic B-splines with 17 control points. Finite-size corrections are needed when including system sizes $20<L<200$. In order to avoid them, we only use system sizes larger than 200 obtaining the critical values 
\begin{equation}\label{eq:pcnu}
 \nu=0.918(5),\quad p_{\rm c}=0.156193(14)\quad.
\end{equation}
Both are compatible with the rough estimates obtained above. 
A careful analysis of finite-size effects does not seem to suggest any drift in system sizes $L>200$. Also, including smaller systems does not seem to spoil the apparent scaling collapse (see system size $L=100$). 
The error bars are obtained using a bootstrap procedure \cite{newman1999monte}, we provide here two times the standard deviation.

\begin{figure}[ht]
 \includegraphics[width=\linewidth]{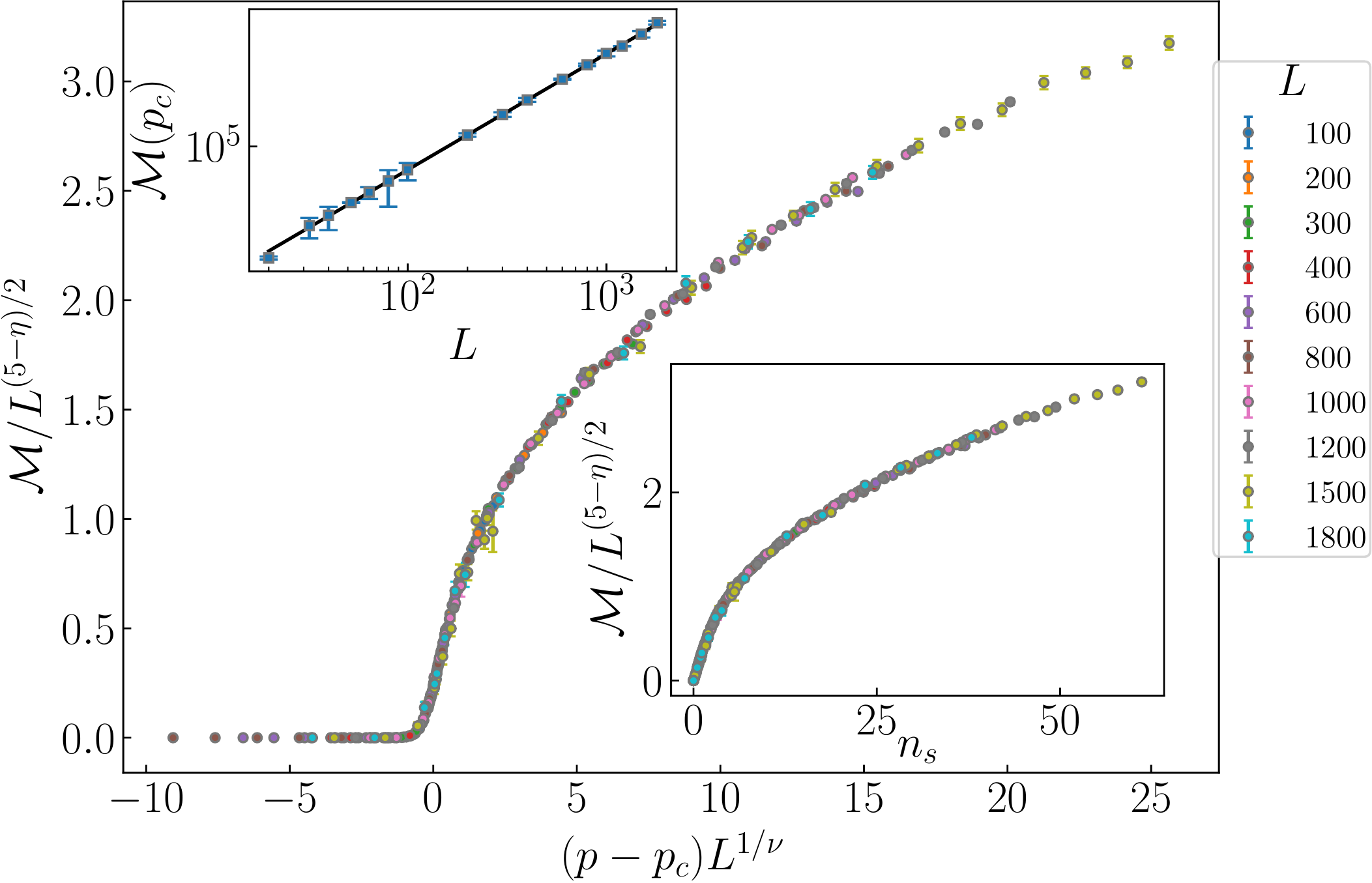}
 \caption{$\mathbf{n=1,\,p=q}$. Scaling collapse of the order parameter $\mathcal{M}/L^{(5-\eta)/2}$ as a function of $x=L^{1/\nu}(p-p_c)$. Upper inset: Value of $M$ at the critical point as a function of the system size. Black dashed line represents a power-law fit to system sizes $L>200$ with exponent $d_f\approx 2.5$. Lower inset: Scaling collapse of the order parameter as a function of $n_s$.}
 \label{fig:Os}
\end{figure}

We can also extract the other independent exponent of this universality class by looking at several other parameters, e.g. the order parameter, the susceptibility or the fractal dimension of the loops. Here, we focus on the number of links of spanning strands, $\M$, a quantity that scales as the order parameter $Q$ \cite{nahum2013phase,nahum2015emergent}. Near the critical point, it follows the scaling law $\M/L^{3}=L^{-\beta/\nu}f_\M(n_s)$, and $\beta/\nu$ is related to the anomalous dimension by the scaling relation $\beta/\nu=1/2+\eta/2$. (Note that it can also be related to the fractal dimension $\M\sim n_s L^{d_f}$ with $d_f=(5-\eta)/2$ \cite{nahum2013phase}, upper inset of Fig.~\ref{fig:Os}). We construct cubic B-splines with 7 control points and obtain the critical exponent $\beta/\nu$. 
The scaling collapse of the curves of the order parameter as a function of $n_s$ (lower inset of Fig.~\ref{fig:Os}) provides an estimate of the anomalous dimension, 
\begin{equation}
 \eta = -0.091(9)\,.
\end{equation}
We use this estimate of $\eta$ and the critical values from Eq.~\ref{eq:pcnu} to plot the scaling collapse of $\M L^{-(5-\eta)/2}$ as a function of $(p-p_c)L^{1/\nu}$ in the main panel of Fig~\ref{fig:Os}. A direct approach gives compatible results (data not shown). Note that $\eta = -0.091(9)$ implies a fractal dimension $d_f=2.546(5)$ or a $\beta/\nu=(1+\eta)/2=0.455(5)$.

\subsection{Line $p=0.1$}
\label{sec:p01}

\begin{figure}[ht]
 \includegraphics[width=\linewidth]{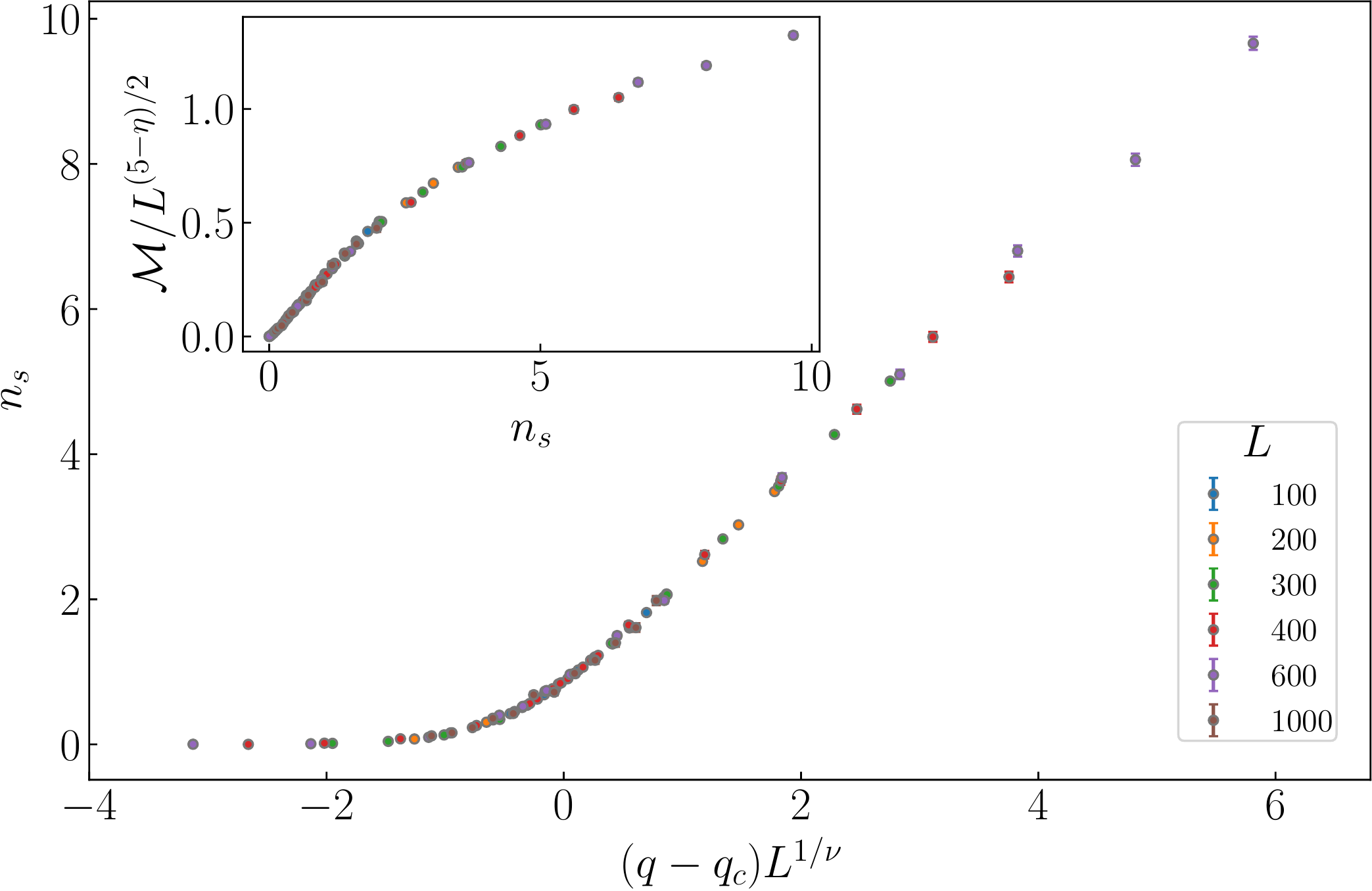}
 \caption{$\mathbf{n=1,\,p=0.1}$. Scaling collapse of $n_s$ as a function of $(q-q_c)L^{1/\nu}$ for the line $p=0.1$. Upper inset: scaling collapse of the order parameter $\mathcal{M}L^{(5-\eta)/2}$ as a function of $n_s$. 
 }
 \label{fig:ns b}
\end{figure}

We have also studied the phase transition in another point of the critical line in order to check universality. Fixing $p=0.1$ and varying $q$, we found that the critical value is at $q_{\rm c}\approx 0.20915$. Figure~\ref{fig:ns b} shows the scaling collapse of the spanning number near this point. The data were fitted to cubic B-splines with 7 nodes. The extracted values of the critical parameters are
\begin{equation}
 \nu_B = 0.911(16),\quad q_{\rm c} = 0.209157(18)\;.
\end{equation}
The upper inset of Fig.~\ref{fig:ns b} shows the scaling collapse of the order parameter for this transition, using cubic B-splines with 6 nodes. The critical exponent obtained is $\eta=-0.07(3)$.

We have extracted the critical value of the spanning number for the two critical points. 
The extrapolated values are compatible and roughly $n_s^{*}=0.97$. 
This is a universal value dependent only on the universality class.

\subsection{Fugacities $n>1$}
\label{sec:ngt1}
We have explored the phase diagrams for integer values of the loop fugacity $n>1$, for small values of $n$. 

The topology of the phase diagram for $n=2$ is very similar to $n=1$, Fig.~\ref{fig:phasediagram}. On the boundaries, there is a finite region with a phase with long loops and two critical points separating it from the short loop phases. These phase transition points are in the universality class of $SU(2)$ quantum magnets in the square lattice \cite{nahum2013phase}. In the interior of the phase diagram, the symmetry of $\CP^1$ is reduced to the one of $\RP^1$, and the transition line is expected to show the universality class of the phase transition of $XY$ magnets in three dimensions.

\begin{figure}[ht]
 \includegraphics[width=1\linewidth]{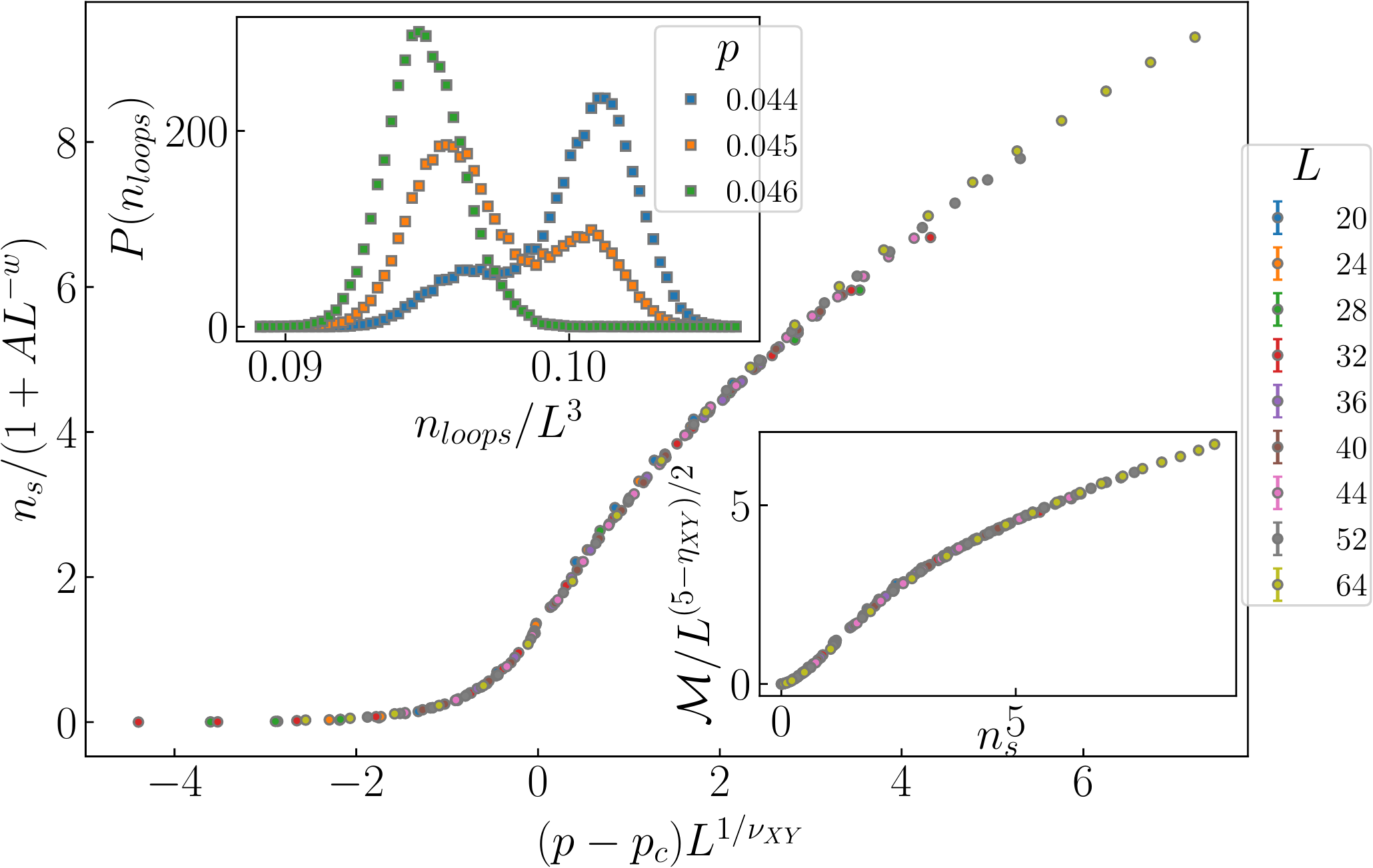}
 \caption{Main panel: $\mathbf{n=2}$ - scaling collapse of $n_s$,  $n_s/(1+AL^{-w_{XY}})$, as a function of $(p-p_c)L^{1/\nu_{XY}}$ with $\nu_{XY}=0.67155$. Fitted parameters are $p_c=0.22523(2)$ and $A=7.4(1)$. Lower inset: scaling collapse $\mathcal{M}/L^{(5-\eta_{XY})/2}$   as a function of $n_s$ without free parameters.  Upper inset: $\mathbf{n=5}$ - distribution of number of loops for size $L=48$, for several values of $p$ close to the transition $p_c=0.045$.  }
 \label{fig:ngt1}
\end{figure}

The main panel of Fig.~\ref{fig:ngt1} shows a test of the universality class of the 3-dimensional $XY$ model for the interior of $n=2$, at the line $p=q$. 
It features the scaling collapse of the spanning number when including simple finite-size corrections, $n_s/(1+AL^{-w_{XY}})$, and as a function of $x=(p-p_c)L^{1/\nu_{XY}}$, given $\nu_{XY}=0.67155$ and $w_{XY}=1.78$ \cite{campostrini2001critical,hasenbusch2019monte}. Note that the only two free parameters used in this scaling collapse are $p_c$ and $A$. To fit these two paremeters, we use cubic B-splines with 11 nodes.

We also test the universality class using the other free critical exponent $\eta_{XY} = 0.038$ \cite{campostrini2001critical,hasenbusch2019monte}, by collapsing $\mathcal{M}/L^{(5-\eta_{XY})/2}$ as a function of the spanning number in the lower inset of Fig.~\ref{fig:ngt1}. Note that there are no free parameters here and there is no need for finite-size corrections. 

A reasonably good scaling collapse can be achieved without including finite-size corrections (data not shown) using values of $\nu\approx0.58$, but an analysis on the finite-size effects shows that this value drifts towards the value of the XY universality class.

For values of $n=3$ and $4$, the boundary lines between short and long loop phases become first-order, but the topology of the phase diagram does not change much (Fig.~\ref{fig:phdiagramsalln}). A more detailed analysis of $n=3$ will be discussed in a different work in the context of first-order transitions \cite{n3_weaklyfirstorder}.

Another interesting case is the line $q=1/2$ and $p>0$, for $n\ge 5$. The upper inset of Fig.~\ref{fig:ngt1} shows the distribution of the number of loops (related to the entropy) very close to the transition point for $n=5$. The characteristic two-peaked shape of the distribution indicates a first-order transition.  The features of a first-order transition can be seen in several other quantities: $n_s$, binder parameter of the number of nodes with crossings, $n_p$, etc.  (Note that even if the energy is related to $n_p$, $n_q$ and $n_{1-q}$, at $q=1/2$ the lattice has a higher symmetry and at $p=0$ the lattice has 4 different equivalent states with symmetry $q\leftrightarrow 1-q$, rendering $n_q$ and $n_{1-q}$ useless for this purpose; the distribution of $n_p$ does show a broadening but with these system sizes the two-peaks are not visible.)


\subsection{Watermelon correlators}
The loop model provides a framework where other relevant features of the $\RP^{n-1}$ in the replica limit $n\to1$ field theory can be computed. An important correlator is the so-called $k$-leg watermelon correlators $G_k(x,x') = \langle O_k(x)O_k(x')\rangle$ \cite{duplantier1989two}. This correlator is a two-point function correlator of the $k$-leg operator 
\begin{equation}
O_k(x) \propto S^1(x)\cdots S^k(x)\quad,
\end{equation}
where $S^i$ are the operator of the $i$-th component of $S$. For $n$ small, supersymmetry can be used to avoid the replica trick \cite{nahum2013loop}. In the loop model, the watermelon correlator can be related to the probability that two sites at a distance $r$ are connected by $k$ strands of loops. Due to the lattice definition and the loop model constraints, it is easier to focus on the 2 and 4-leg watermelon correlators, i.e., the two sites are connected by 2 or 4 strands.  Far from a critical point, these correlators decay exponentially to 0 in the disordered phase, or to a constant in the long loop phase. At a critical point they follow a power-law behaviour $G_k(r)\propto r^{-2 x_k}$, where $x_k$ is its scaling dimensions. In particular, 
the scaling dimension $x_2$ is related to the anomalous dimension by $x_2= (1+\eta)/2$ \cite{nahum2013loop}. Analogously to \cite{hasenbusch2011anisotropic}, we focus on the integral of these operators by computing
\begin{equation}
  \langle C_2\rangle = \frac{1}{V}\sum_{x} C_2(0,x),\, \langle C_4\rangle = \frac{1}{V}\sum_{x} C_4(0,x)\,,
\end{equation}
where the sums run through all the nodes in the lattice, except the origin, and $V=3 L^3/4-1$ is the number of terms of the sum. $C_2(0,x)$ is 1 if the nodes at 0 and at $x$ are connected by at least two strands of a loop, 0 otherwise. $C_4(0,x)$ is 1 if the nodes at 0 and at $x$ are connected by 4 strands but they do not belong to the same loop, 0 otherwise. 
These quantities are proportional to the integral of the watermelon correlators, and at the critical point, they go to 0 following a power-law $L^{-2x_k}$.

\begin{figure}[ht]
 \includegraphics[width=0.95\linewidth]{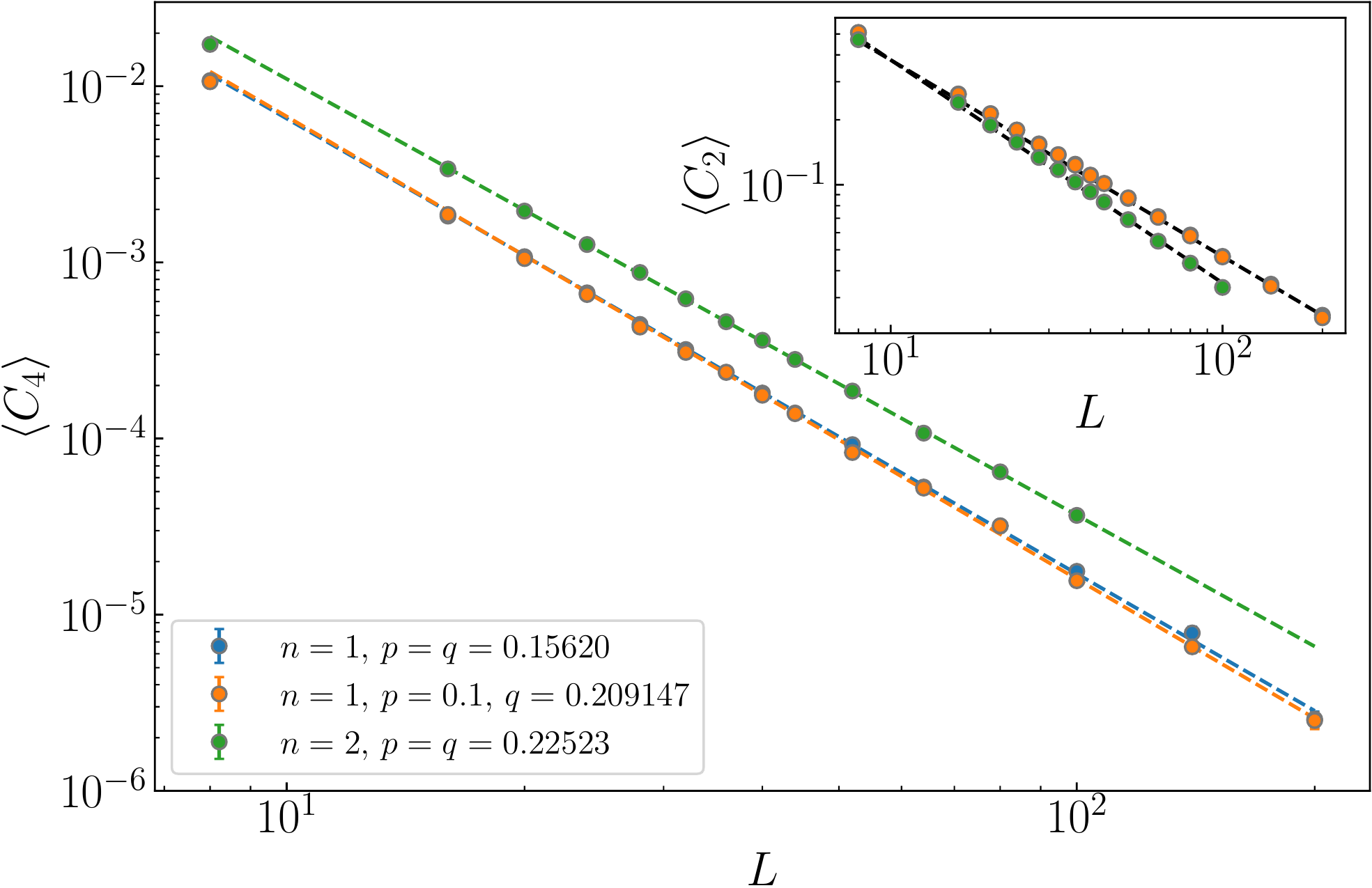}
 \caption{Main panel: 4-leg watermelon correlators $\langle C_4 \rangle$ as a function of the system size $L$, for $n=1$ at $p=q=0.15620$ (blue) and $p=0.1,\,q=0.209147$ (orange) and for $n=2$ at $p=q=0.22523$. Each straight line is a fit to a power-law with exponent $-2x_4$, with values $x_4=1.282$ (blue), $x_4=1.302$ (orange) and $x_4=1.235$ (green). Inset: 2-leg watermelon correlators $\langle C_2\rangle$ as a function of the system size $L$, same color code. Straight lines represent power-laws with scaling dimension obtained from scaling collapse for $n=1$ ($x_2=0.455$) and from the literature $n=2$ ($x_2=0.517$).}
 \label{fig:watermelon}
\end{figure}

Inset of Fig.~\ref{fig:watermelon} shows $\langle C_2\rangle$ as a function of the system size $L$, both for $(p,q) =(0.15624,0.15624)$ and $(0.1,0.22523)$ for $n=1$ and $(0.22523,0.22523)$ for $n=2$. Fits to power-laws give the values 
$x_2=0.467(2)$, $0.473(2)$ and $0.536(2)$
, respectively. 
They both seem to be compatible with the values of the anomalous dimension obtained by a scaling collapse when including sizes $L\le100$ (data not shown), as finite-size effects are present for $\eta$. 
Power-laws using $\eta$ from $n=1$ and $n=2$ are shown as black dashed lines, corresponding to $x_2=0.455$ ($\eta=-0.09$) and $0.517$ ($\eta_{XY}=0.038$) respectively for comparison. 
A small finite-size effect is present, as it can be observed by the difference in values of the fitted exponents.

The main panel of Fig.~\ref{fig:watermelon} shows $\langle C_4\rangle$ as a function of the system size $L$. Fits to power-laws give 
$x_4 = 1.282(12)$ and $1.302(12)$, for $n=1$ lines $p=q$ and $p=0.1$ respectively. They are both compatible with each other and the average is then,
\begin{equation}
x_4 = 1.292(8)\quad.
\end{equation} 
For $n=2$, the fitted scaling dimension is $x_4=1.235(4)$, which can be compared to the scaling dimension from the spin-2 operator anisotropic perturbation, $1.2361(11)$ \cite{hasenbusch2011anisotropic}, or $\Delta_t$ from conformal bootstrap, $1.23629(11)$ \cite{chester2020carving}. 
A finite-size effect analysis does not provide significantly different values for any of these scaling dimensions in this range of system sizes. However, from the results of the 2-leg watermelon correlator, we can expect them to be not negligible.

The values of the 4-leg watermelon scaling dimension for the CPLC are significantly different from those of the loop models without crossings, which corresponds to $\CP^{n-1}$ sigma models. Particularly for $n=1$ where preliminary data shows $x_4\approx1.20$.



\section{Conclusions}

We have characterized a new universality class, estimating the main critical exponents with high precision in two different points of the boundary line. 
Both sets of values are compatible with each other, in line with the assumption that the whole boundary line is indeed in the same universality class. 
On the other hand, this set of values is different from the class of Anderson transition with symplectic symmetry or froom any of the usual classes \cite{asada2005anderson, slevin2014critical}, suggesting this is indeed a new universality class. 
Coincidentally, this value is similar to the exponent found in a transition between a Dirac semimetal and a metal in a disordered system  with $\nu\approx 0.92$ \cite{kobayashi2014density}. Although the properties of the phases are clearly different, whether this is a coincidence or not could be explored in future works. 

We have also explored the phase diagram for integer values of $n>1$, and we have shown that within this space of parameters, on the line $q=1/2$ for integer values of $n$ there is no multicritical point or any continuous transition. 
We have shown that $\RP^{n-1}$ is a good description of the CPLC. This is particularly interesting for $n=3$, where it maps to the smectic-isotropic transition. We will study this in detail in a forthcoming work.


This loop model can be easily modified to provide access to a critical point related to the deconfined critical points in 2+1D XY magnets \cite{qin2017duality,zhang2018continuous}. 
There has been a lot of work already in this system, and whether the character of the transition is first-order or continuous is not completely resolved. Numerical results are compatible with a theoretical framework with a very slow RG flow close to a fixed point with emergent $O(4)$ symmetry \cite{zhao2019symmetry,serna2019emergence, ma2020theory, nahum2020note}, although other possibilities are not discarded. 
A framework where non-integer values of $n$ could be simulated would be very interesting as well.

Finally, from the soft matter point of view, this loop model and its realizations could be modified by introducing topological constraints.
This would be related to very hard problems for polymer physics, where field theory descriptions are still missing \cite{de1979scaling,halverson2014melt,imakaev2015effects,serna2015topological}. 
This framework seems to be ideal to implement these constraints and explore this mysterious territory.

\begin{acknowledgments}
I thank Adam Nahum, Andr\'es Somoza, Miguel Ortuño and John Chalker for very useful discussions and collaborations on loop models. Special thanks to Adam Nahum and Andr\'es Somoza for insightful comments about the manuscript and for encouraging me to finish this work. I want to thank Jose Juan Fernandez-Melgarejo for useful discussions. 
This work was funded by Fundaci\'on S\'eneca grant 19907/GERM/15.
\end{acknowledgments}


\appendix
\section{Samples}
\label{app:samples}

\begin{figure}[ht]
  \includegraphics[width=0.99\linewidth]{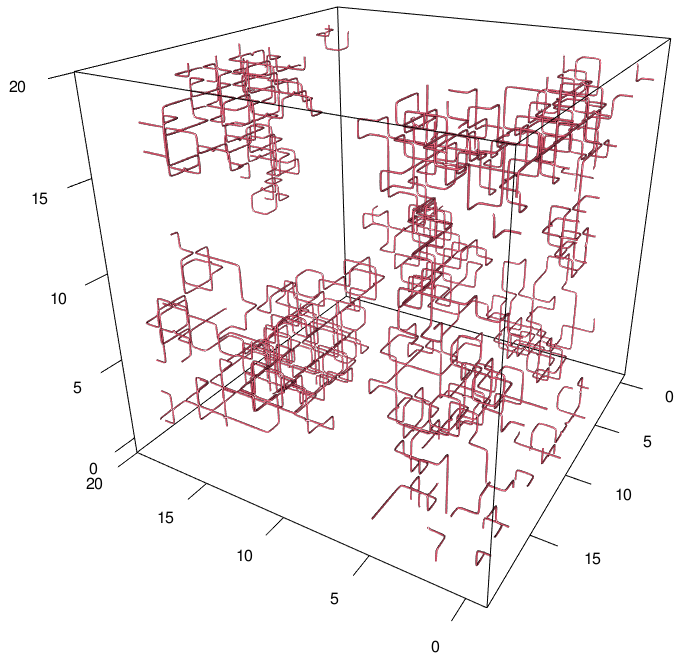}
 \caption{Example of a fractal unoriented loop with periodic boundary conditions in a system size of $L=20$ at $p=q=0.1562$}
 \label{fig:samples long loops}
\end{figure}

At $n=1$, where $p_c=q_c\approx0.1562$ loops are fractal and span the whole sample. An example of this fractal loop with $d_f\approx 2.55$ is shown in Fig.~\ref{fig:samples long loops} for a system of size $L=20$ with periodic boundary conditions.

\section{Phase diagrams}

\begin{figure}[ht]
 \includegraphics[width=0.48\linewidth]{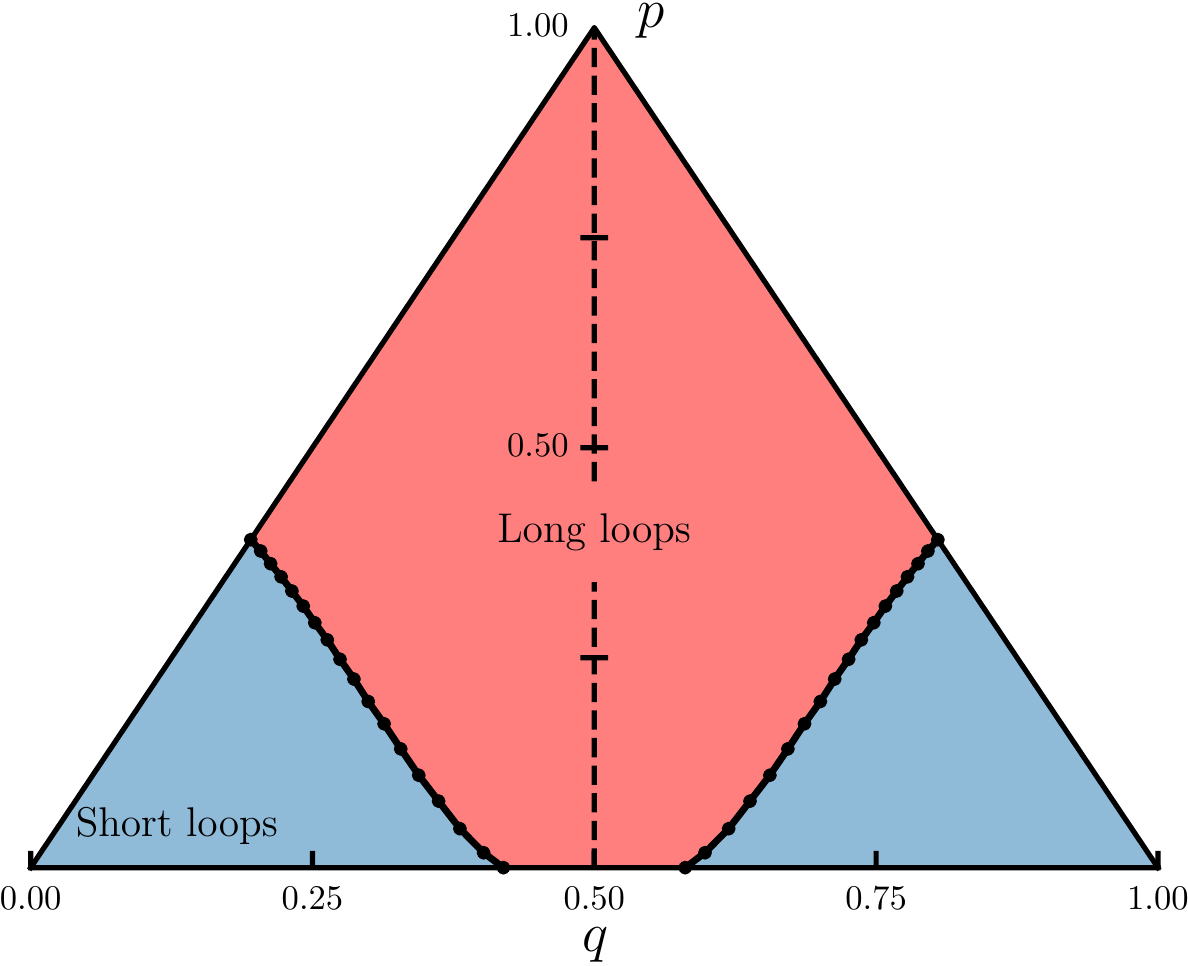}
 \includegraphics[width=0.48\linewidth]{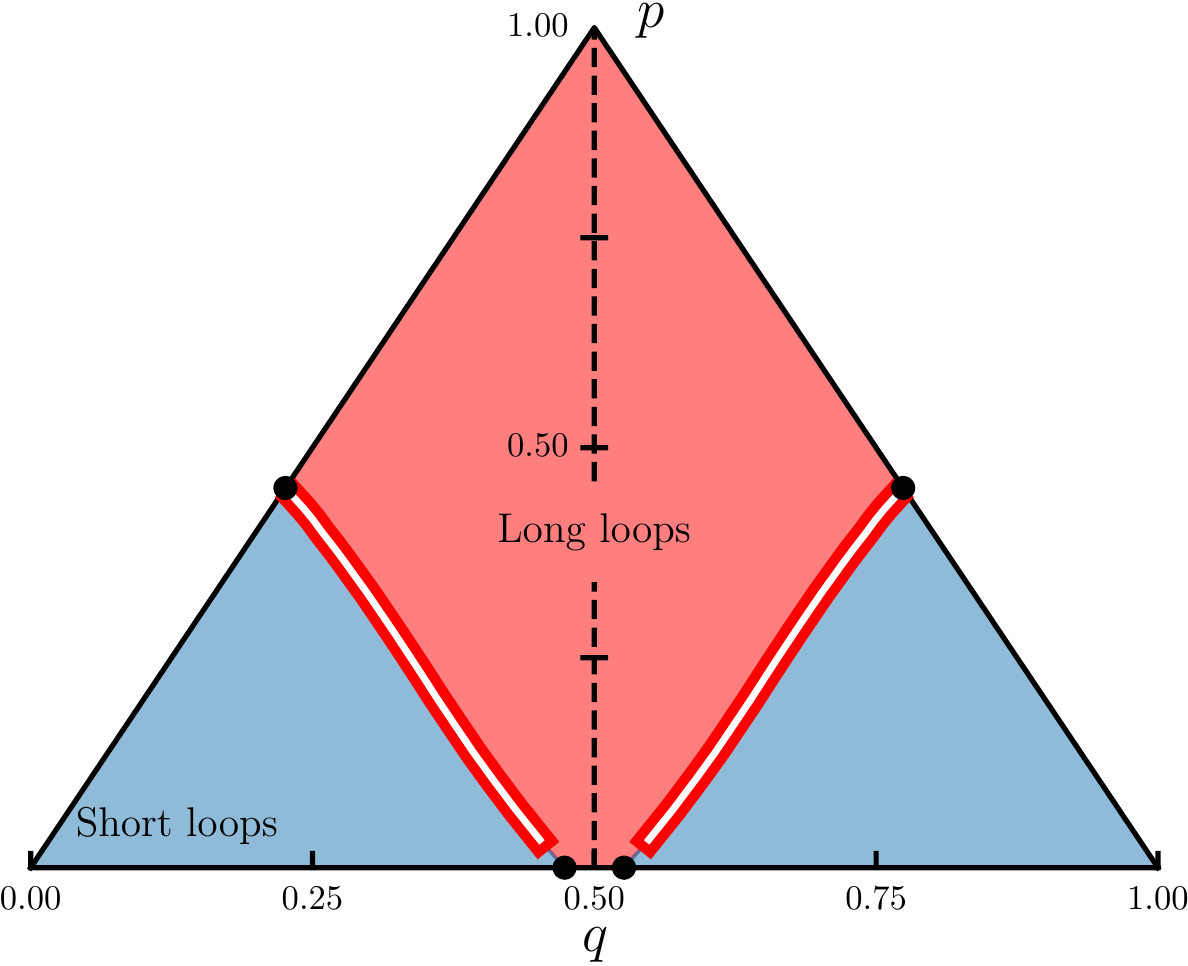}
 \includegraphics[width=0.48\linewidth]{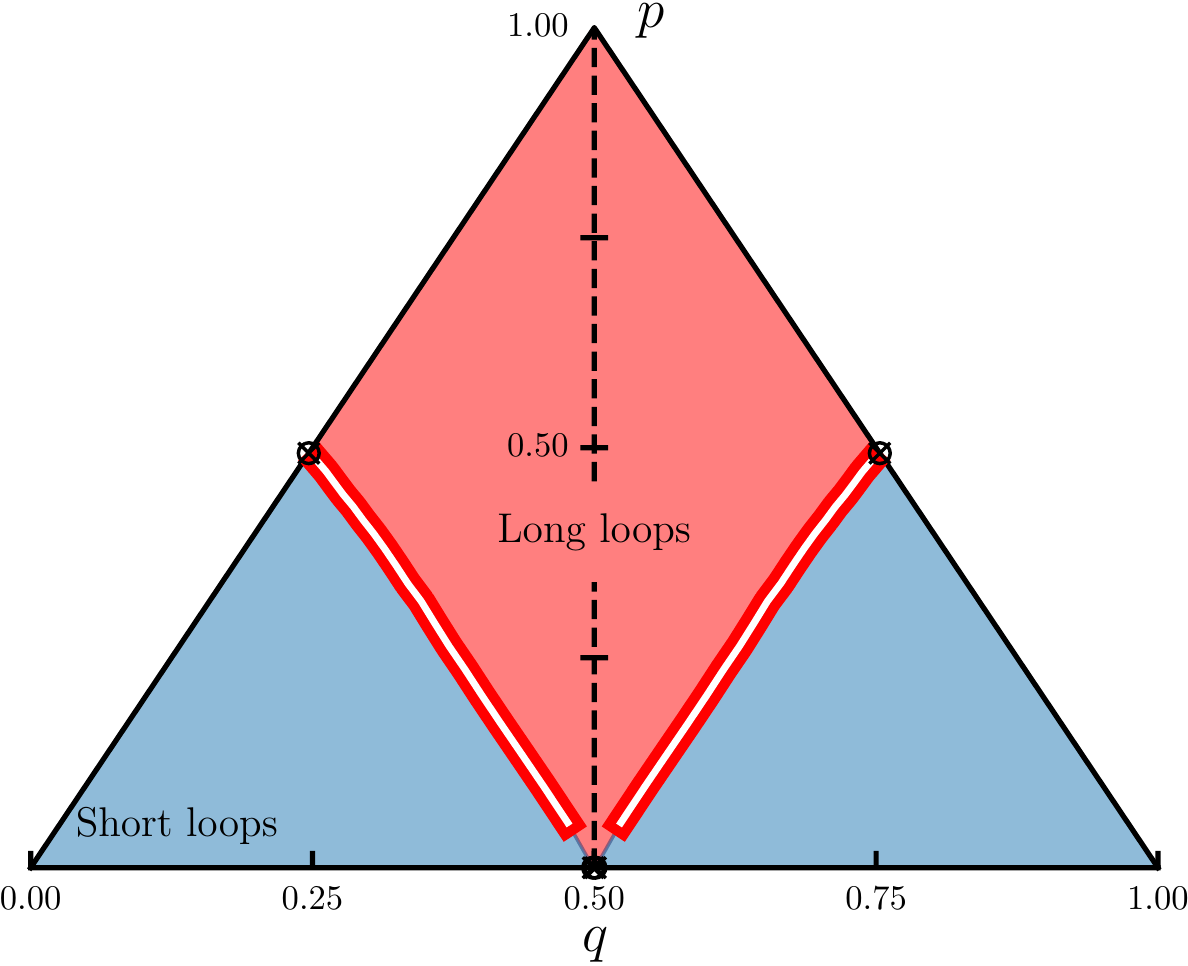}
 \includegraphics[width=0.48\linewidth]{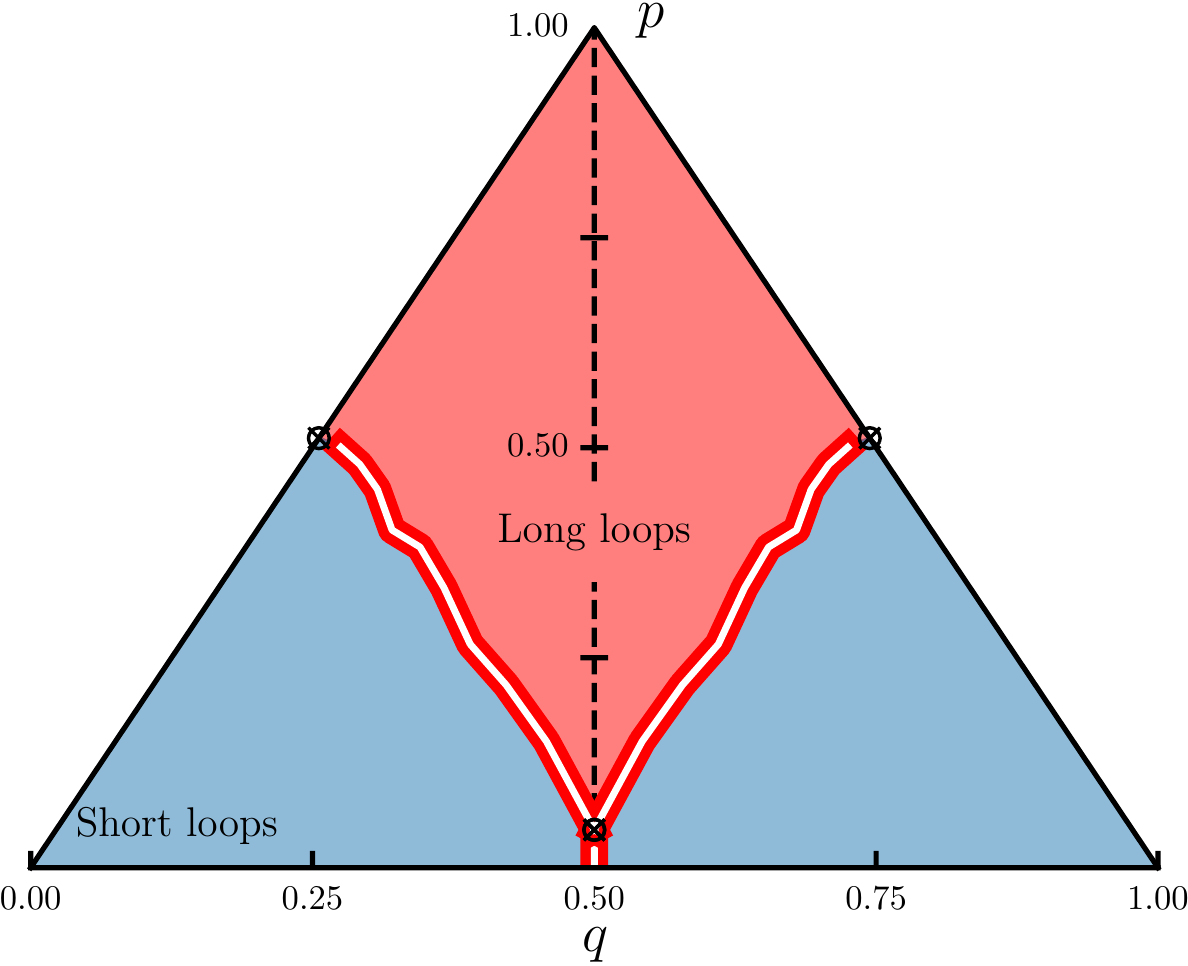}
 \caption{Phase diagrams for $n =$ 2, 3, 4 and 5. The phase with long loops (red) is separated from the phases with short loops by a boundary line that is continuous for $n=2$ (black line) and first-order for the rest (red and white line). For $n=5$ there is a first-order transition between the two phases with short loops.}
 \label{fig:phdiagramsalln}
\end{figure}

Phase diagram of the CPLC for values of $n=$2, 3, 4 and 5 is shown in Fig.~\ref{fig:phdiagramsalln}. These boundary lines have been estimated from $n_s$ crossings for small system sizes ($L=$16 and 32), except for the points in black, where we have used much larger sizes and a more detailed study.

\bibliography{Unoriented3d.bib}

\end{document}